\documentclass[%
 reprint,
 amsmath,amssymb,
 aps,
 pra,
]{revtex4-2}

\usepackage{graphicx}
\usepackage{dcolumn}
\usepackage{bm}
\usepackage{hyperref}
\begin{document}

\preprint{APS/123-QED}

\title{Rigorous theory of thin-vapor-layer linear optical properties:\\The case of quenching of atomic polarization \\ upon collisions of atoms with dielectric walls}

\author{A. V. Ermolaev}
\author{T. A. Vartanyan}
\affiliation{
 ITMO University, Kronversky Prospect 49, 197101 St. Petersburg, Russia
}
\begin{abstract}
Hot alkali metal vapors enclosed in sub-micron spectroscopic cells provide an ideal system for fundamental studies of the atom-wall and atom-light interactions at nanoscale. Here, we propose a novel approach for calculating the eigenmodes of a thin-vapor-layer beyond the limitations of the first-order perturbation theory in optical density for the case of quenching of atomic polarization upon collisions of atoms with dielectric walls. We show that higher-order optical density corrections lead to a remarkable density-dependent blue shift and deformation of the spectral line shapes of reflection, transmission, and absorption. We also demonstrate that the eigenmodes of the thin-vapor-layer can be calculated independently of the choice of optical boundary conditions. This greatly extends the applicability of the constructed theory for the development of miniature atomic sensors.
\end{abstract}

\maketitle

\section{\label{sec:level1}Introduction}
Recently, there has been a growing interest in fundamental and applied studies on light interaction with isolated atoms, since the processes occurring in such systems are the most accessible to theoretical description. Newly developed methods for cooling and trapping atoms have allowed to realize (with a high degree of approximation) an ideal atomic system that does not interact with anything. Unfortunately, these techniques remain expensive and cumbersome and the time, during which light can interact with such systems, is substantially limited~\cite{KITCHING}. A much cheaper and more available alternative is hot atomic vapors, which can be enclosed in miniature spectroscopic cells and used continuously for extended periods of time~\cite{SARKISYAN2001,PhysRevApplied.14.034054}. Studying the optical properties of thin-vapor-layers (TVL) are highly relevant due to a number of important applications, such as practical implementation of atomic sensors, magnetometers and methods for frequency stabilization of lasers, as well as the possibility to conduct fundamental research into the nature of interaction of atoms with dielectric surfaces, detailed studies of collision processes in atomic ensembles, and effects leading to the shift and broadening of spectral lines~\cite{ADAMS,Ritter,CHIP,URIEL:21,PhysRevA.97.053841,PhysRevLett.123.173203}.

The manifestation of Doppler-free structures in the spectra of light reflection from the interface of dielectric medium and atomic vapor was first demonstrated by Cojan~\cite{refId0} and was followed by the series of remarkable experiments with sodium vapor~\cite{WOERDMAN1975248,burgmans:jpa-00208462}. In particular, Cojan pointed out the inconsistency of the conventional dispersion theory, based on the local relationship between the field and the induced polarization in the gaseous medium, when describing the observed phenomenon. Following Cojan's ideas, the exact solution to the problem of light reflection from a semi-infinite layer of resonant gas was subsequently obtained by Schuurmans~\cite{schuurmans:jpa-00208442}. As was noted by Cojan and Schuurmans, the main reason leading to the narrowing of spectral lines is the transient process of establishing polarization after the collision of an atom with a wall. Indeed, regardless of the nature of atom-surface interactions, immediately after the collision, the atom "sees" a driven field with the different detuning compared to the one possessed before the collision due to the thermal motion. In a dilute gas, the mean free path of an atom without coherence loss may become greater than the wavelength of the incident light. Consequently, the strong influence of the spatial dispersion induced by both the thermal motion of atoms and boundary presence leads to the formation of the sub-Doppler structure in reflection spectra. 

Further studies of resonant reflection [also known as selective reflection (SR)] of light from the boundary of a gaseous medium also included the processes of non-linear nature~\cite{VART1985,DUCLOY1988}. Besides, SR can be used to develop methods for narrowing the line of laser generation~\cite{VART1990}, in high-resolution studies~\cite{SINGH1986107} and in detailed consideration of the dynamics of atom collisions with a dielectric wall. Moreover, a number of publications consider the influence of higher-order effects in vapor density~\cite{GUO1994}, antireflection coatings~\cite{VARTANYAN1994}, and the Lorentz-Lorenz field correction~\cite{GUO1996}. In Ref.~\cite{BLUESHIFT} the paradoxical blue shift of the resonant frequency was studied in the reflection spectrum (hereinafter referred to as "blueshift"); it was revealed that the blueshift is sensitive to the concentration of atomic vapors. This phenomenon was previously prescribed to the transient polarization aspect in~\cite{schuurmans:jpa-00208442,GUO1996}, however, without explaining in detail its origin.

Brand new prospects appeared after the theoretical prediction of the possibility to enhance the optical response of a resonant vapor spatially confined between two dielectric media in a layer with a thickness of the order of the incident wavelength~\cite{PhysRevA.51.1959,ZAMBON1997308} and practical implementation of miniature vapor cells containing vapors of alkali metals~\cite{SARKISYAN2001}. The works cited subsequently gave a rise to several remarkable studies of the Paschen-Back and related effects on the hyperfine structure of alkali metals~\cite{Sargsyan:12,Sargsyan:17,Sargsyan:17_2,Sargsyan:2014}; atom-surface interactions at nanoscale~\cite{ADAMS2, ADAMS,KEAVENEY}; interactions of the resonant atomic ensemble with the plasmonic structures~\cite{LEVY,Uriel} and other applications~\cite{APP1,APP2,Ritter,CHIP}. Nevertheless, the underlying theoretical model~\cite{PhysRevA.51.1959} had such drawbacks as the exclusion of light reflection from the rear gas-dielectric interface and the limitation by the first order of the perturbation theory (PT) in vapor density. The first challenge was tackled in~\cite{Dutier:03}, where authors took into account the interplay between the field confined inside the Fabry–Pérot (FP) resonator and selective contribution from the vapor. This revisited approach has been used in numerous studies aimed at interpreting experimental observations, while remaining limited to only the first order of PT.

As can be seen, research in the field of spectroscopy of thin layers of hot atomic vapors in the previous twenty years has been actively expanding to newly applied and fundamental fields. However, the theoretical description is still incomplete for various physical processes, which occur during the resonant interaction of light with vapor spatially limited at nanoscale. The existing models describe qualitatively the optical response of a highly rarefied gaseous medium, although the underlying approximations can lead to significant discrepancy between theory and experiment. From this point of view, in addition to experimental studies, theoretical works intended to clarify, expand, and revise already existing models are also of great interest. The purpose of our work is to construct the universal solutions to the TVL problem beyond the scope of the first-order PT and to study the effect of these higher-order contributions on the line shape, width, shift of the maxima and other features of the reflection, transmission, and absorption spectra of the TVL spatially-confined between transparent dielectric media. 

The paper is organized as follows: Sec.~\ref{sec:Theoretical} describes the underlying assumptions of our model, considered geometry, and the Maxwell-Bloch set of equations along with the boundary conditions that fully describe the self-consistent TVL problem. In Sec.~\ref{sec:Methods} for the first time we introduce the iterative PT approach that serves to find the eigenmodes of the TVL in the prescribed order with respect to the optical density of atomic vapor. The calculation results of reflectivity, transmittivity and absorptivity of a TVL, their dependence on the system parameters are given in Sec.~\ref{sec:Results}. Finally, in Sec.~\ref{sec:Discussion}, we discuss the peculiarities associated with the higher-order vapor density corrections for the cases of specular and quenching atom-wall collisions and focus our attention on the blueshift phenomenon arising in the higher-order optical density solutions.

\section{\label{sec:Theoretical}Theoretical background}
\begin{figure}[b]
\includegraphics{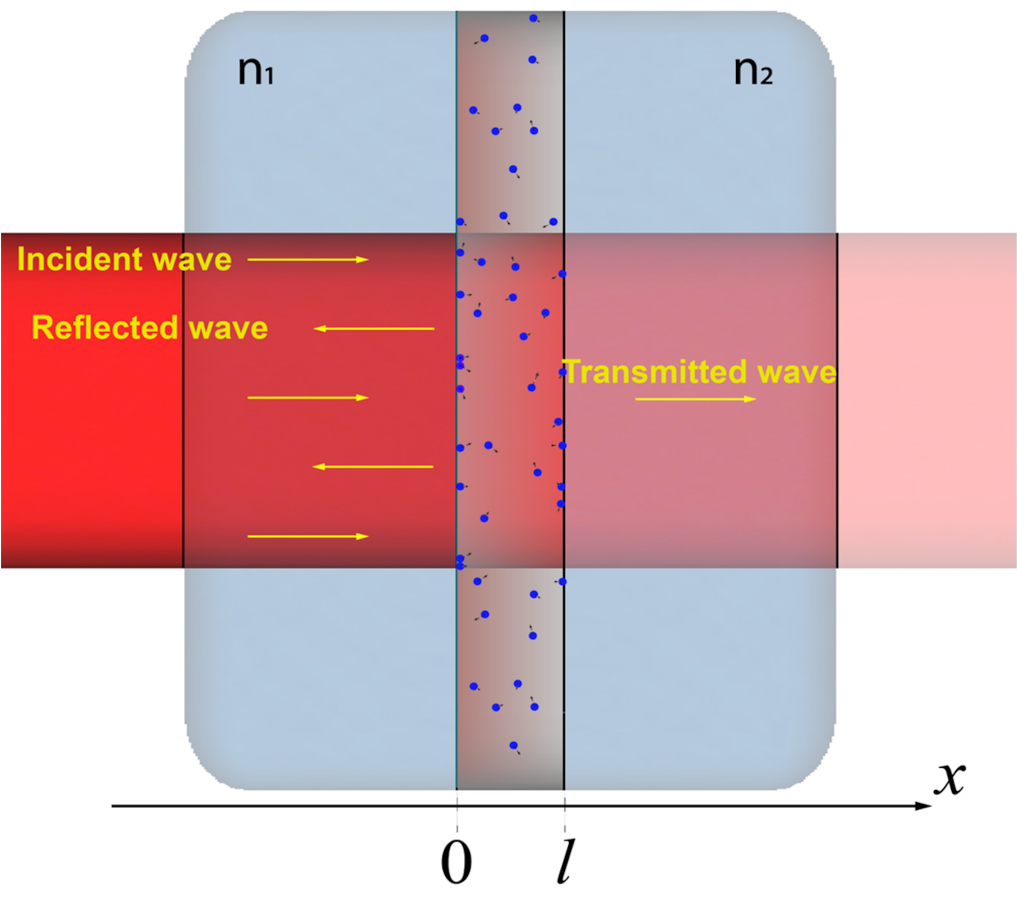}
\caption{\label{fig:description}Schematic illustration of the reflection and transmission of light through a TVL confined between two dielectric media in the region $0 \leq x \leq l$.}
\end{figure}

Consider the resonant light interaction with the atomic vapor spatially confined between two transparent dielectric media that are taken to be semi-infinite. Inclusion of the effects connected with the presence of the outer boundaries of these media is trivial and will not be treated here. In Fig.~\ref{fig:description} we schematically represent the one-dimensional (1D) geometry of the considered problem, where $l$ denotes the thickness of the gas layer, while $n_{1}$ and $n_{2}$ stand for the refractive indices of the surrounding media. 

\noindent - The vapor layer consists of two-level atoms with $\omega_{0}$ being the transition frequency between the ground and excited states;

\noindent - The normally incident laser radiation could be considered as a linearly-polarized plane monochromatic electromagnetic wave with a frequency $\omega$ varying in the spectral vicinity of the resonant frequency, i.e. $|\omega - \omega_{0}|/\omega \ll 1$. Hence, the use of rotating wave approximation is justified;

\noindent - We restrict ourselves to the linear regime of interactions in which the incident power is so low, that it could not saturate the resonant transition. Moreover, we do not take into account the optical pumping effect;

Finally, throughout our consideration, we treat the above problem using the semi-classical approach. In the following section we introduce the exact form of the Maxwell-Bloch set of equations in the weak driven field limit that fully takes into account the effect of strong spatial dispersion~\cite{VART2001,PhysRevA.101.053850}.

\subsection{Maxwell-Bloch system of equations in the linear regime of interactions}

We begin by writing down the wave equation obtained from the microscopic Maxwell's equations in non-magnetic media in the absence of free charges and current using the Gaussian convention

\begin{equation}
\Delta \bm{E}=\frac{1}{{{c}^{2}}}\frac{{{\partial }^{2}}}{\partial {{t}^{2}}}\left( \bm{E}+4\pi \bm{P} \right),
\label{eq:wave-vector}
\end{equation}
where $\bm{E}$ and $\bm{P}$ are electric field and polarization vectors, respectively, and $c$ is the speed of light. In accordance with the considered 1D problem, the above equation reduces to a scalar

\begin{equation}
\frac{{{d}^{2}}E(x)}{d{{x}^{2}}}+{{k}^{2}}E(x)=-4\pi {{k}^{2}}P(x),
\label{eq:wave-scalar}
\end{equation}
where $k = 2\pi/\lambda$. In the above equation we expressed field and polarization within the gaseous medium in the following form neglecting nonlinear optical processes
\begin{equation}
E(x,t)=\frac{1}{2}E(x)\exp (-i\omega t)+c.c.,
\label{eq:field}
\end{equation}
\begin{equation}
P(x,t)=\frac{1}{2}P(x)\exp (-i\omega t)+c.c..
\label{eq:polarization}
\end{equation}
In the density matrix formalism for a two-level system, the macroscopic polarization $P(x)$ could be expressed in terms of the off-diagonal density matrix element $\rho_{21}(x,\upsilon,t)=\rho_{21}(x,\upsilon)\exp (-i\omega t)$ as
\begin{equation}
P(x)=2Nd\int_{-\infty }^{+\infty }{f(\upsilon )\rho_{21} \left( x,\upsilon  \right)d\upsilon},
\label{eq:macroscopic}
\end{equation}
where $d$ denotes the transition dipole moment, $f(\upsilon)$ stands for the 1D velocity distribution of atoms inside the gas layer, and $\upsilon$ being the projection of atomic velocity onto the $x$ axis. The most reasonable assumption for $f(\upsilon)$ at room temperatures is the Maxwell-Boltzmann distribution function \cite{Todorov,Todorovproc}, which we use in further calculations
\begin{equation}
f(\upsilon) = (\sqrt{\pi}\upsilon_{T})^{-1} \exp({-\upsilon^2/\upsilon_{T}^2}),
\label{eq:Max-Bol}
\end{equation}
where $\upsilon_{T} = \sqrt{2k_{B}T/M}$ denotes the most probable thermal velocity.

The density matrix 
\begin{equation}
\hat{\rho }=\left( \begin{matrix}
   {{\rho }_{11}}\left( x,\upsilon  \right) & \rho_{12} \left( x,\upsilon  \right)  \\
   {{\rho }_{21}}\left( x,\upsilon  \right) & {{\rho }_{22}}\left( x,\upsilon  \right)  \\
\end{matrix} \right)
\label{eq:density matrix}
\end{equation}
satisfies the Liouville–von Neumann equation
\begin{equation}
i\hbar \frac{d}{dt}\hat{\rho }=\left[ \hat{H_{0}} + \hat{V},\hat{\rho } \right]-\frac{i\hbar}{2} {{\left[ \hat{\Gamma },\hat{\rho } \right]}_{+}},
\label{eq:Liouville}
\end{equation}
here $[\hat{a},\hat{b}]=\hat{a}\hat{b}-\hat{b}\hat{a}$ and $[\hat{a},\hat{b}]_{+}=\hat{a}\hat{b}+\hat{b}\hat{a}$ denote commutator and anti-commutator, respectively, $\rho_{11,22} \left( x,\upsilon  \right)$ describes the population of the ground and excited state (diagonal elements of the density matrix satisfy the normalization conditions $\rho_{11}+\rho_{22}=1$), $\hat{\Gamma}$ is the relaxation matrix,
$\hat{H_{0}}$ the Hamiltonian of the undisturbed system, and $\hat{V} = -\hat{d}E(x)$ ($\hat{d}$ is the dipole moment operator).
Taking into account that
\[\frac{d}{dt}\rho_{21} (x,\upsilon ,t)=\left( \frac{\partial }{\partial t}+\upsilon \frac{\partial }{\partial x} \right)\rho_{21} (x,\upsilon )\exp \left( -i\omega t \right),\]
from Eq.~(\ref{eq:Liouville}) we obtain the equation of motion for the off-diagonal density matrix element

\begin{equation}
\upsilon \frac{\partial \rho_{21} (x,\upsilon )}{\partial x}+\left( \gamma +i\Delta  \right)\rho_{21} (x,\upsilon )=\frac{id}{2\hbar }E(x),
\label{eq:off-dioganal}
\end{equation}
where $\Delta=\omega_{0}-\omega$ and $\gamma$ denotes the homogeneous width of the spectral line, i.e. the sum of natural and collisional widths. Equations~(\ref{eq:wave-scalar}),~(\ref{eq:macroscopic}),~(\ref{eq:Max-Bol}), and~(\ref{eq:off-dioganal}) form a self-consistent set of equations governing the dynamics of the consideration system in the linear regime of interactions.

\subsection{Dimensionless variables}
Before proceeding to the solution of the above system of integro-differential equations, for the sake of simplicity we rewrite them in a convenient form by introducing the dimensionless variables: 
\begin{subequations}
\label{eq:whole}
\begin{eqnarray}
m = \frac{2\sqrt{\pi}Nd^{2}}{\hbar k \upsilon_{T}},
\label{eq:m}
\end{eqnarray}
\begin{eqnarray}
\xi = kx,
\label{eq:ksi}
\end{eqnarray}
\begin{eqnarray}
\nu = \upsilon/\upsilon_{T},
\label{eq:nu}
\end{eqnarray}
\begin{eqnarray}
\Gamma = \gamma/k\upsilon_{T},
\label{eq:Gamma}
\end{eqnarray}
\begin{eqnarray}
\Omega = \Delta/k\upsilon_{T},
\label{eq:Omega}
\end{eqnarray}
\begin{eqnarray}
\eta = \Gamma-i\Omega,
\label{eq:eta}
\end{eqnarray}
and
\begin{eqnarray}
\sigma(\xi,\nu) = \frac{2\hbar k \upsilon_{T}}{id}\rho_{21}(\xi,\nu).
\label{eq:eta}
\end{eqnarray}
\end{subequations}
With the following choice of dimensionless variables, the Maxwell-Bloch system of equations that fully accounts for the non-local optical response can be written as 

\begin{equation}
\frac{{{d}^{2}}E(\xi )}{d{{\xi }^{2}}}+E(\xi )=-2im\int_{-\infty }^{+\infty }{\sigma (\xi ,\nu )\exp \left( -{{\nu }^{2}} \right)d\nu },
\label{eq:field-dim1}
\end{equation}
\begin{equation}
\nu \frac{\partial \sigma (\xi ,\nu )}{\partial \xi }+\eta \sigma (\xi ,\nu )=E(\xi ).
\label{eq:sigma-dim1}
\end{equation}
In order to consider accurately the structure of the field inside the gas layer, the set of Eqs.~(\ref{eq:field-dim1})-(\ref{eq:sigma-dim1}) should be solved self-consistently with the particular choice of boundary conditions for the field and off-diagonal element of the density matrix at two gas-dielectric medium interfaces situated at $\xi=0$ and $\xi=kl=\phi$.

\subsection{Boundary conditions}
In accordance with the geometry of the problem, the incident, reflected and transmitted fields could be written in the following way
\begin{eqnarray*}
  & E_{in}\exp({in_{1}\xi}), \\ 
 & E_{r}\exp({-in_{1}\xi}), \\
 & E_{t}\exp\left[{in_{2}(\xi - \phi)}\right], 
\label{eq:inc}
\end{eqnarray*}
respectively. Then, the continuity conditions at glass-vapor interfaces impose that
\begin{subequations}
\begin{eqnarray}
E_{in}+E_{r}=E(0),
\end{eqnarray}
\begin{eqnarray}
in_{1}\left( E_{in}-E_{r} \right)=E'(0),
\end{eqnarray}
\begin{eqnarray}
E_{t}=E(\phi),
\end{eqnarray}
\begin{eqnarray}
in_{2}E_{t}=E'(\phi),
\end{eqnarray}
\label{eq:BC}
\end{subequations}
where $(')$ stands for the derivative with respect to $\xi$. Basically, the above boundary conditions serve to single out the unique solutions for the field within the vapor layer. The form of Eqs.~(\ref{eq:BC}) is dictated by the continuity of the tangential components of electric and magnetic fields at both vapor boundaries. 
The choice of the appropriate boundary conditions for $\sigma(\xi,\nu)$ depends on the nature of atom-wall collisions. At the highest level of generality, the polarization of the atoms that leave the surface with any particular velocity is related to the polarizations of the atoms arriving at the surface with all different velocities in the ensemble. In lieu of the comprehensive theoretical as well as experimental results on this complicated problem it is common to consider two limiting cases of the interactions of atoms with the surface of a dielectric material: specular reflection of atoms with the polarization preserved and diffuse scattering of atoms with the polarization quenching.  Rigorous theory of TVL linear optical properties for the case of specular atom-wall collisions was presented in our previous paper~\cite{PhysRevA.101.053850}, where we showed that the given set of Eqs.~(\ref{eq:field-dim1})-(\ref{eq:sigma-dim1}) can be solved explicitly by means of Fourier series expansion of the field inside the gaseous medium. Here we consider another limiting case of atom-dielectric surface interactions in which the electron excitation, and thus the induced polarization are lost upon the collision. This assumption is based on the numerous studies on the alkali-metal atoms interactions with the surface of dielectric material (see, for example, Ref.~\cite{Zajonc,Duclou1993}). At room temperatures, these collisions are usually governed by the adsorption and desorption processes, in which the angular distribution of atoms outgoing the dielectric interface may be approximated using Knudsen cosine or related laws. Formally, in the framework of the considered problem we could write
\begin{subequations}
\begin{eqnarray}
\sigma(\xi=0,\nu>0) = 0,
\label{eq:sigma_grat}
\end{eqnarray}
and
\begin{eqnarray}
\sigma(\xi=\phi,\nu<0) = 0,
\label{eq:sigma_less}
\end{eqnarray}
\end{subequations}
for the polarization at the first and second boundaries, respectively. Boundary conditions~(\ref{eq:sigma_grat})-(\ref{eq:sigma_less}) complete the formulation of the considered problem.

\section{\label{sec:Methods}Methods}
In this section we derive step by step the universal solution of the self-consistent system of equations for the field and polarization inside the gaseous medium. First of all, we demonstrate that the initial system of two equations for $E(\xi)$ and $\sigma(\xi,\nu)$ [see Eqs.~(\ref{eq:field-dim1}) and~(\ref{eq:sigma-dim1})] in the linear regime of interactions along with the diffuse boundary conditions Eqs.~(\ref{eq:sigma_grat}) and~(\ref{eq:sigma_less}) could be converted to one integro-differential equation of the Fredholm type. Then, we introduce in details the iterative PT method allowing one to compute the eigenmodes of the TVL semi-numerically in any order of PT with respect to the atomic number density. Finally, by imposing the continuity conditions [see Eq.~(\ref{eq:BC})], we write down the exact expressions for the reflection and transmission of a TVL. In the course of our consideration, we rely on a rigorous mathematical method for solving differential equations on the interval of continuity of the coefficients with boundary conditions given in different regions of space, described in details in Ref.~\cite{SMIRNOV}.
\subsection{Self-consistent field equation}
The general solution of Eq.~(\ref{eq:sigma-dim1}) reads as
\begin{equation}
\sigma \left( \xi ,\nu  \right)=\exp \left( -{\eta \xi }/{\nu } \right)\left[ C+{{\nu }^{-1}}\int_{0}^{\xi}{E\left( \xi'  \right)\exp \left( {\eta \xi' }/{\nu} \right)d\xi' } \right],
\label{eq:solution}
\end{equation}
where the arbitrary constant $C$ is to be determined from the boundary conditions Eqs.~(\ref{eq:sigma_grat})-(\ref{eq:sigma_less}). The particular solutions take different forms for atoms moving in opposite directions with velocities $\nu>0$ and $\nu<0$ 
\begin{equation}
\sigma \left( \xi ,\nu >0 \right)={{\nu }^{-1}}\int_{0}^{\xi }{E\left( {{\xi }'} \right)\exp \left[ {\eta \left( {\xi }'-\xi  \right)}/{\nu } \right]d{\xi }'},
\label{eq:sig+1}
\end{equation}
and
\begin{equation}
\sigma \left( \xi ,\nu <0 \right)={{\nu }^{-1}}\int_{\phi }^{\xi }{E\left( {{\xi }'} \right)\exp \left[ {\eta \left( {\xi }'-\xi  \right)}/{\nu } \right]d{\xi }'},
\label{eq:sig-1}
\end{equation}
respectively. It is important to emphasize that in above equations we do not make any assumptions regarding the structure of the field inside the gaseous medium. 

Now, we could write down Eq.~(\ref{eq:field-dim1}) in the following form
\begin{widetext}
\begin{eqnarray}
	\frac{{{d}^{2}}E\left( \xi  \right)}{d{{\xi }^{2}}}+E\left( \xi  \right)=  -2 i m \Bigg[ \int_{-\infty }^{0}{\sigma \left( \xi ,\nu <0 \right)\exp \left( -{{\nu }^{2}} \right)d\nu }
	+\int_{0}^{\infty }{\sigma \left( \xi ,\nu >0 \right)\exp \left( -{{\nu }^{2}} \right)d\nu }\Bigg] .
\label{eq:E}
\end{eqnarray}
Now, by combining Eqs.~(\ref{eq:sig+1}-\ref{eq:E}) together, we get
\begin{eqnarray}
\frac{{{d}^{2}}E\left( \xi  \right)}{d{{\xi }^{2}}}+E\left( \xi  \right)=m \Bigg[\int_{-\infty }^{0}{\int_{\phi }^{\xi }{E\left( {{\xi }'} \right)K\left( \xi -{\xi }',\nu  \right)d{\xi }'}d\nu }  +\int_{0}^{\infty }{\int_{0}^{\xi }{E\left( {{\xi }'} \right)K\left( \xi -{\xi }',\nu  \right)d{\xi }'}d\nu }\Bigg],
\label{eq:E1}
\end{eqnarray}
\end{widetext}
where we have introduced the integral kernel
\begin{equation}
K\left( \xi, \nu \right)=-2i \frac{\exp \left( -{{\nu }^{2}}{-\eta \xi }/{\nu } \right)}{\nu }.
\label{eq:kernel}
\end{equation}
Finally, taking into account that 
\begin{equation}
K\left( \left| \xi -{\xi }'\right|, \nu  \right)=\left\{ \begin{matrix}
   K\left( \xi -{\xi }', \nu \right),\text{ }\xi \ge {\xi }'  \\
   K\left( {\xi }'-\xi, \nu  \right),\text{ }\xi <{\xi }'  \\
\end{matrix}, \right.
\label{eq:KERNEL}
\end{equation}
we can rewrite the obtained equation in the following compact way
\begin{eqnarray}
\frac{{{d}^{2}}E\left( \xi  \right)}{d{{\xi }^{2}}}+E\left( \xi  \right) = m \nonumber \\ \times \int_{0}^{\infty}\int_{0}^{\phi }{E\left( {{\xi }'} \right)K\left( \left| \xi -{\xi }' \right|, \nu \right) d{\xi }' d\nu}.
\label{eq:E2}
\end{eqnarray}

From Eq.~(\ref{eq:E2}) we can directly see that we are dealing with a self-consistent field problem. Equations of this type were studied extensively in the past. For instance, in the case of a thick gas layer (i.e. for $\phi \to \infty$) the obtained integro-differential equation was solved via the Fourier transform method and the Wiener-Hopf technique for the specular and quenching boundary conditions, respectively~\cite{schuurmans:jpa-00208442}. The exact solution for the case of quenching collisions is still absent and the extension of the Wiener-Hopf method to the case of a finite layer thickness seems too cumbersome~\cite{Baraff}. At the same time, the existing solutions of this problem are limited only to the first order of PT, and their generalization to higher orders of magnitude in optical density remains unclear~\cite{Dutier:03,PhysRevA.51.1959,ZAMBON1997308}. In what follows we are focused on the construction of the universal solution for the TVL problem for the case of quenching boundary conditions in the framework of the higher-order PT and on studying the effects arising from these higher-order contributions.

\subsection{Converting the integro-differential field equation to an integral equation}
We start our consideration by defining two linearly-independent solutions of Eq.~(\ref{eq:E2}). Let $E(\xi)$ be the solution of the above self-consistent equation. It can be easily verified that $E(\phi - \xi)$ is another linearly-independent solution. As the vapor slice with plane parallel boundaries possesses mirror symmetry with respect to the plane going through the center of gaseous medium $\xi = \phi/2$ it is convenient to use a set of linearly-independent solutions which have definite properties with respect to reflection in this plane, namely, the even and odd solutions of Eq.~(\ref{eq:E2}). For an even (symmetric) solution the boundary conditions correspond to equality of values of the function itself and opposite signs of its derivative on the edges of the interval, and vice versa for an odd (antisymmetric) solution - to the opposite signs of the function itself and equality of its derivatives on the edges of the interval. 
Then, the solution of the left-hand side of Eq.~(\ref{eq:E2})
\begin{equation}
\frac{{{d}^{2}}E_{0}\left( \xi  \right)}{d{{\xi }^{2}}}+E_{0}\left( \xi  \right)=0.
\label{eq:homog}
\end{equation}
could be written as
\begin{equation}
{{E}_{0}}\left( \xi  \right)={{c}_{1}}\cos \left( \xi -{\phi}/{2}\right)+{{c}_{2}}\sin \left( \xi -{\phi}/{2}\right).
\label{eq:homog_sol}
\end{equation}
Following the standard procedure, we construct the system of equations for the variative constants $c_{1}$ and $c_{2}$
\begin{eqnarray}
\left[ \begin{matrix}
   \cos \left( \xi -{\phi }/{2} \right) & \sin \left( \xi -{\phi }/{2} \right)  \\
   -\sin \left( \xi -{\phi }/{2} \right) & \cos \left( \xi -{\phi }/{2} \right)  \\
\end{matrix} \right] \left[ \begin{matrix} {{c}_{1}}^{\prime }\left( \xi  \right) \\ {{c}_{2}}^{\prime }\left( \xi  \right) \end{matrix} \right] = \nonumber\\
\left[ \begin{matrix}
   0  \\
   m\int_{0}^{\infty }\int_{0}^{\phi }{E\left( {{\xi }'} \right)K\left( \left| \xi -{\xi }' \right|, \nu \right)d{\xi }' d\nu}  \\
\end{matrix} \right]
\label{eq:Wronsk}
\end{eqnarray}
Assuming the right-hand of the Eq.~(\ref{eq:E2}) to be known, its solution could be formally written in terms of the Green's function with the use of Eq.~(\ref{eq:Wronsk}) as
\begin{widetext}
\begin{equation}
E\left( \xi  \right)= C_{1} \cos(\xi - \phi/2) +C_{2}\sin(\xi - \phi/2) + m\int_{0}^{\infty }d\nu\int_{{\phi }/{2}}^{\xi }{\sin \left( \xi - {\xi }'  \right) d{\xi }' \int_{0}^{\phi }{E\left( {{\xi }''} \right)K\left( \left| {\xi }'-{\xi }'' \right|, \nu \right)d{\xi }''}},
\label{eq:GREENl}
\end{equation}
where $C_{1}$ and $C_{2}$ are arbitrary constants determined by the choice of linearly-independent solutions. Requiring the symmetry conditions for the even solution introduced above, we obtain $C_{2} = 0$, while $C_{1}$ can be chosen arbitrarily. Therefore, for the even branch of Eq.~(\ref{eq:GREENl}) one could write down

\begin{equation}
E_{e}\left( \xi  \right)=\cos(\xi - \phi/2) + m\int_{0}^{\infty } d\nu \int_{{\phi }/{2}}^{\xi }{\sin \left( \xi - {\xi }' \right) d{\xi }' \int_{0}^{\phi }{E_{e}\left( {{\xi }''} \right)K\left( \left| {\xi }'-{\xi }'' \right|, \nu \right)d{\xi }''}},
\label{eq:GREEN_even}
\end{equation}
where we set $C_{1}=1$. Similarly, for an odd solution by choosing $C_{1}=0$ and $C_{2}=1$ we obtain
\begin{equation}
E_{o}\left( \xi  \right)=\sin(\xi - \phi/2) + m\int_{0}^{\infty} d\nu \int_{{\phi }/{2}}^{\xi }{\sin \left( \xi - {\xi }' \right) d{\xi }' \int_{0}^{\phi }{E_{o}\left( {{\xi }''} \right)K\left( \left| {\xi }'-{\xi }'' \right|, \nu \right)d{\xi }''}}.
\label{eq:GREEN_odd}
\end{equation}
\end{widetext}

\subsection{Iterative PT method}
Eqs.~(\ref{eq:GREEN_even}) and~(\ref{eq:GREEN_odd}) could be solved iteratively by means of the series expansion of the field with respect to the optical density $m$. First of all, we represent the even and odd solutions as the following perturbation series
\begin{equation}
E_{e/o}\left( \xi  \right)={{E}_{e/o}^{(0)}}\left( \xi  \right)+m{{E}_{e/o}^{(1)}}\left( \xi  \right)+{{m}^{2}}{{E}_{e/o}^{(2)}}\left( \xi  \right)+\ldots,
\label{eq:series}
\end{equation}
where the superscript denotes the order of PT. After that, we substitute the form of the fields~(\ref{eq:series}) into the Eqs.~(\ref{eq:GREEN_even})-(\ref{eq:GREEN_odd}), and equate the terms at the same powers of $m$. From Eqs.~(\ref{eq:GREEN_even})-(\ref{eq:series}) one could directly find the recurrent equation
\begin{eqnarray}
E_{e/o}^{(n)}\left( \xi  \right)=\int_{0}^{\infty }d{\nu}\int_{{\phi }/{2}}^{\xi }{d{\xi }'\sin \left( {\xi }-{\xi}'  \right)} \nonumber \\ \times \int_{0}^{\phi }{E_{e/o}^{(n-1)}}\left( {{\xi }''} \right)K\left( \left| {\xi }'-{\xi }'' \right|, \nu \right)d{\xi }'',
\label{eq:iteration}
\end{eqnarray}
It is worth noting that the convergence of Eq.~(\ref{eq:series}) is guaranteed by the smallness of optical density paremeter, which is usually of the order $m=2\sqrt{\pi}Nd^{2}/\hbar k \upsilon_{T} \ll 1$ under typical experimental conditions with alkali metal vapors.

\subsection{Reflectivity and transmittivity of the TVL}
Equations~(\ref{eq:GREEN_even})-(\ref{eq:iteration}) allow us to find the solution of the initial set of Maxwell-Bloch equations for a field in the region of space $0 \leq \xi \leq \phi$ up to the prescribed order of PT. Now, in order to calculate the reflection and transmission coefficients of the TVL one has to imply the dielectric boundary conditions [See Eq.~(\ref{eq:BC})] on the obtained solution. Setting the amplitude of the incident wave to be unit, we can rewrite the above continuity conditions in the following way
\begin{subequations}
\begin{eqnarray}
1+r=aE_{e}\left( 0 \right)+bE_{o}\left( 0 \right),
\end{eqnarray}
\begin{eqnarray}
i{{n}_{1}}\left( 1-r \right)=a{{\left. \left( {dE_{e}}/{d\xi } \right) \right|}_{\xi =0}}+b{{\left. \left( {dE_{o}}/{d\xi } \right) \right|}_{\xi =0}},  
\end{eqnarray}
\begin{eqnarray}
t=aE_{e}\left( \phi  \right)+bE_{o}\left( \phi  \right),
\end{eqnarray}
\begin{eqnarray}
i{{n}_{2}}t=a{{\left. \left( {dE_{e}}/{d\xi } \right) \right|}_{\xi =\phi }}+b{{\left. \left( {dE_{o}}/{d\xi } \right) \right|}_{\xi =\phi}},
\end{eqnarray}
\label{eq:SYSTEM1}
\end{subequations}

where $r$ and $t$ are the amplitude reflection and transmission coefficients, respectively, while  $a$ and $b$ are the coefficients of the linear set of equations. Taking into account the symmetry relations of the even and odd fields and their derivatives at the boundaries below we write down the exact solution of the set of Eqs.~(\ref{eq:SYSTEM1}) with respect to $r$ and $t$ calculated up to the prescribed $n$-th order of PT in atomic number density
\begin{equation}
r=\frac{\left( {{n}_{1}}-{{n}_{2}} \right)\left( I_{1}I_{4}+I_{2}I_{3}\right)+2i\left( {{n}_{1}}{{n}_{2}}I_{1}I_{2}+I_{3}I_{4} \right)}{\left( {{n}_{1}}+{{n}_{2}} \right)\left( I_{1}I_{4}+I_{2}I_{3} \right)+2i\left( {{n}_{1}}{{n}_{2}}I_{1}I_{2}-I_{3}I_{4} \right)},
\label{eq:r}
\end{equation}
\begin{equation}
t=\frac{2{{n}_{1}}(I_{1}I_{4}-I_{2}I_{3})}{\left( {{n}_{1}}+{{n}_{2}} \right)\left( I_{1}I_{4}+I_{2}I_{3} \right)+2i\left( {{n}_{1}}{{n}_{2}}I_{1}I_{2}-I_{3}I_{4} \right)},
\label{eq:t}
\end{equation}
where we introduced for convenience
\begin{align}
\begin{split}
 & I_{1}= E_{e}|_{\xi =0} = E_{e}|_{\xi =\phi}, \\ 
 & I_{2}= E_{o}|_{\xi =0} = - E_{o}|_{\xi =\phi}, \\ 
 & I_{3}=\frac{\partial }{\partial \xi }E_{e}|_{\xi =0} = -\frac{\partial }{\partial \xi }E_{e}|_{\xi =\phi}, \\ 
 & I_{4}=\frac{\partial }{\partial \xi } E_{o}|_{\xi =0} = \frac{\partial }{\partial \xi } E_{o}|_{\xi =\phi}. \\ 
\end{split}
\label{eq:str}
\end{align}
Finally, in accordance with the described PT approach Eqs.~(\ref{eq:r})-(\ref{eq:str}) could be calculated with the accuracy up to the prescribed order $n$ with respect to the optical density. In this case, the above terms have the following structure
\begin{align*}
\begin{split}
 & I_{1}^{(n)}={{\left[ E_{e}^{(0)}+mE_{e}^{(1)}+{{m}^{2}}E_{e}^{(2)}+...+{{m}^{n}}E_{e}^{(n)} \right]}_{\xi =0}}, \\ 
 & I_{2}^{(n)}={{\left[ E_{o}^{(0)}+mE_{o}^{(1)}+{{m}^{2}}E_{o}^{(2)}+...+{{m}^{n}}E_{o}^{(n)} \right]}_{\xi =0}}, \\ 
 & I_{3}^{(n)}=\frac{\partial }{\partial \xi }{{\left[ E_{e}^{(0)}+mE_{e}^{(1)}+{{m}^{2}}E_{e}^{(2)}+...+{{m}^{n}}E_{e}^{(n)} \right]}_{\xi =0}}, \\ 
 & I_{4}^{(n)}=\frac{\partial }{\partial \xi }{{\left[ E_{o}^{(0)}+mE_{o}^{(1)}+{{m}^{2}}E_{o}^{(2)}+...+{{m}^{n}}E_{o}^{(n)} \right]}_{\xi =0}}. \\ 
\end{split}
\label{eq:str}
\end{align*}
Equations~(\ref{eq:r})-(\ref{eq:str}) are completely general and allow us to calculate the reflectivity and transmittivity of the TVL surrounded by dielectric media with refractive indices $n_{1}$ and $n_{2}$ in the prescribed order of the PT. The particular form of the terms $I_{1}^{(n)}$ - $I_{4}^{(n)}$ in the zeroth, first and second orders of PT discussed in detail in the following sections.

\section{\label{sec:Results}Results}
In this section we proceed to the calculation of reflectivity, transmittivity and absorptivity of TVL following the PT approach introduced in the previous section. At the first stage of the consideration, we search for two linearly-independent field solutions which can be computed with the accuracy determined by the prescribed order of PT using Eq.~(\ref{eq:iteration}). It is important to underline that the eigenmodes of the TVL calculated via this procedure are the unique field solutions which are valid for any environment of the gas layer. In this work we consider the case of dielectric environment and to this end, on the second stage of our calculations we apply the continuity conditions~(\ref{eq:SYSTEM1}) allowing us to directly compute the reflectivity and transmittivity of TVL confined between two dielectric media [See Eqs.~(\ref{eq:r})-(\ref{eq:str})].

\subsection{Zeroth and first-order results}
\label{sec:Zero}
We begin with the step-by-step derivation of the first-order PT solution of the TVL problem. To find an even field solution, we substitute an even zero solution $E_{e}^{(0)}=\cos(\xi-\phi/2)$ into Eq.~(\ref{eq:iteration})
\begin{eqnarray}
E_{e}^{(1)}\left( \xi  \right)=\int_{0}^{\infty }d\nu\int_{{\phi }/{2}}^{\xi }{\sin \left( {\xi }-{\xi}'  \right) d{\xi }'} \nonumber \\ \times\int_{0}^{\phi }{\cos({{\xi}'' -\phi/2}) K\left( \left| {\xi }'-{\xi }'' \right|,\nu \right)d{\xi }'' }.
\label{eq:EVEN1}
\end{eqnarray}
Spatial integration over ${\xi}'$, ${\xi}''$ in Eq.~(\ref{eq:EVEN1}) could be done analytically, below we present only the result of these calculations
\begin{widetext}
\begin{eqnarray}
E_{e}^{(1)}\left( \xi  \right)=-2i\int_{0}^{\infty }{d\nu \frac{\eta }{{{\eta }^{2}}+{{\nu }^{2}}}{{e}^{-{{\nu }^{2}}}}} \Bigg\{ \left( \xi -{\phi }/{2} \right)\sin \left( \xi -\frac{\phi }{2} \right)  \nonumber \\ +\frac{{{\nu }^{2}}}{{{\eta }^{2}}+{{\nu }^{2}}}\left[ {{e}^{-\frac{\xi \eta }{\nu }}}+{{e}^{\frac{\eta \left( \xi -\phi  \right)}{\nu }}}-2{{e}^{-\frac{\eta \phi }{2\nu }}}\cos \left( \xi -\frac{\phi }{2} \right) \right]\left[ \frac{\nu \sin \left( {\phi }/{2} \right)}{\eta }-\cos \left( {\phi }/{2} \right) \right] \Bigg\}.
\label{eq:EVEN1SOL}
\end{eqnarray}
Similar solution could be obtained for the odd branch in the first order of PT by substituting $E_{o}^{(0)}=\sin(\xi-\phi/2)$ into Eq.~(\ref{eq:iteration})
\begin{eqnarray}
E_{o}^{(1)}\left( \xi  \right)=2i\int_{0}^{\infty }{d\nu \frac{\eta }{{{\eta }^{2}}+{{\nu }^{2}}}{{e}^{-{{\nu }^{2}}}}} \Bigg\{ \left( \xi -\phi/2  \right)\cos \left( \xi -\frac{\phi }{2} \right)-\sin \left( \xi -\frac{\phi }{2} \right) \nonumber \\ -\frac{{{\nu }^{2}}}{{{\eta }^{2}}+{{\nu }^{2}}}\left[ {{e}^{-\frac{\xi \eta }{\nu }}}-{{e}^{\frac{\eta \left( \xi -\phi  \right)}{\nu }}}+{2{{e}^{-\frac{\eta \phi }{2\nu }}}\eta \sin \left( \xi -\frac{\phi }{2} \right)}/{\nu } \right] \left[ \frac{\nu \cos \left( {\phi }/{2} \right)}{\eta }+\sin \left( {\phi }/{2} \right) \right] \Bigg\}.
\label{eq:ODD1SOL}
\end{eqnarray}
\end{widetext}
In above Eqs. velocity integration (i.e. the integration over dimensionless parameter $\nu=\upsilon/\upsilon_{T})$ has to be performed numerically. Before this, we define the fields and corresponding derivatives at the boundaries of the layer in accordance with Eq.~(\ref{eq:str}). This could simply be done by evaluating even and odd field solutions [see Eqs.~(\ref{eq:EVEN1SOL})-~(\ref{eq:ODD1SOL})] and their $\xi$ derivatives at $\xi = 0$. The exact form of terms $I_{1}^{(1)}, I_{2}^{(1)}$ and $I_{3}^{(1)}, I_{4}^{(1)}$ is presented in Appendix. 
\begin{figure*}[!htb]
\includegraphics{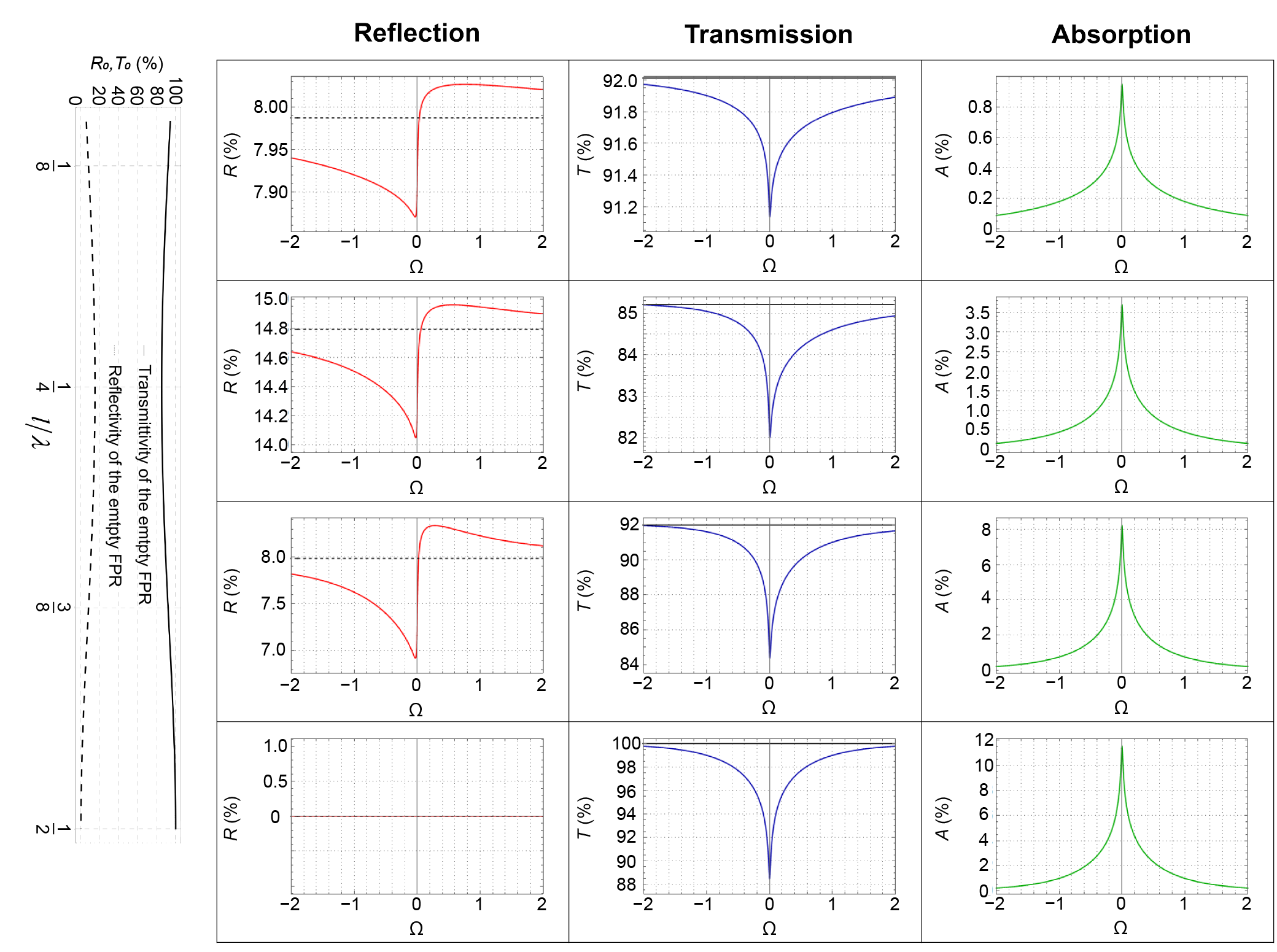}
\caption{\label{fig:first order}Left panel: Reflectivity and transmittivity of the empty FP resonator against the gap thickness between two transparent dielectric media with $n_{1} = n_{2} = 1.5$. Right panel: Reflection, transmission, and absorption spectra of the vapor layer sandwiched between dielectric media at various thicknesses of the gas layer: from $\lambda/8$ to $\lambda/2$ with the step $\lambda/8$ (from top to bottom) for $\gamma/k\upsilon_{T}=0.01$, $m=0.002$. The horizontal dashed (in reflection) and solid (in transmission) black lines correspond to the spectra of the empty FP resonator with the thickness indicated on the left panel. The error of the calculations is of the order of $m^2$ (details of the calculations are highlighted in the text).}
\end{figure*}
The first terms in Eqs.~(\ref{eq:I11})-(\ref{eq:I44}) correspond to the solution for the field in vacuum. By setting $m = 0$ in the given expressions $I_{1}^{(1)}-I_{4}^{(1)}$, and then substituting them into Eqs.~(\ref{eq:r}) and~(\ref{eq:t}), one may directly obtain the well-known expressions for the reflectivity and transmittivity of the empty FP resonator
\begin{equation}
{{R}_{0}}={{\left| {{r}_{0}} \right|}^{2}}={{\left| \frac{i\left( {{n}_{1}}-{{n}_{2}} \right)\cos \phi +\left( {{n}_{1}}{{n}_{2}}-1 \right)\sin \phi }{i\left( {{n}_{1}}+{{n}_{2}} \right)\cos \phi +\left( {{n}_{1}}{{n}_{2}}+1 \right)\sin \phi } \right|}^{2}},
\label{eq:R0}
\end{equation}
\begin{equation}
{{T}_{0}}={{\left| {{t}_{0}} \right|}^{2}}=1-{{R}_{0}}.
\label{eq:T0}
\end{equation}

This zero solution with respect to the density of atomic vapor leads to a $\lambda/2$-periodic dependence of reflection and transmission on the thickness of the gap between dielectric media. On the left panel of Fig.~\ref{fig:first order} we plot $R_{0}$ and $T_{0}$ as a function of dimensionless thickness $\phi=2\pi l/\lambda$, while on the right panel we present the numerical calculation of the reflection, transmission and absorption of the TVL versus the dimensionless detuning $\Omega = \Delta/k\upsilon_{T}$ at the corresponding thicknesses. The absorption spectra were calculated using the energy conservation law
\begin{equation}
R + T + A = 1,
\label{eq:A}
\end{equation}
where $R = |r|^{2}$ and $T = |t|^{2}$. While calculating spectra on the right panel of Fig.~\ref{fig:first order} we kept only the zeroth- and first-order terms with respect to $m$ in Eqs.~(\ref{eq:r})-(\ref{eq:t}). 
The obtained spectra indicate the presence of purely sub-Doppler structures in TVL spectra.
\begin{figure}[!htb]
\includegraphics{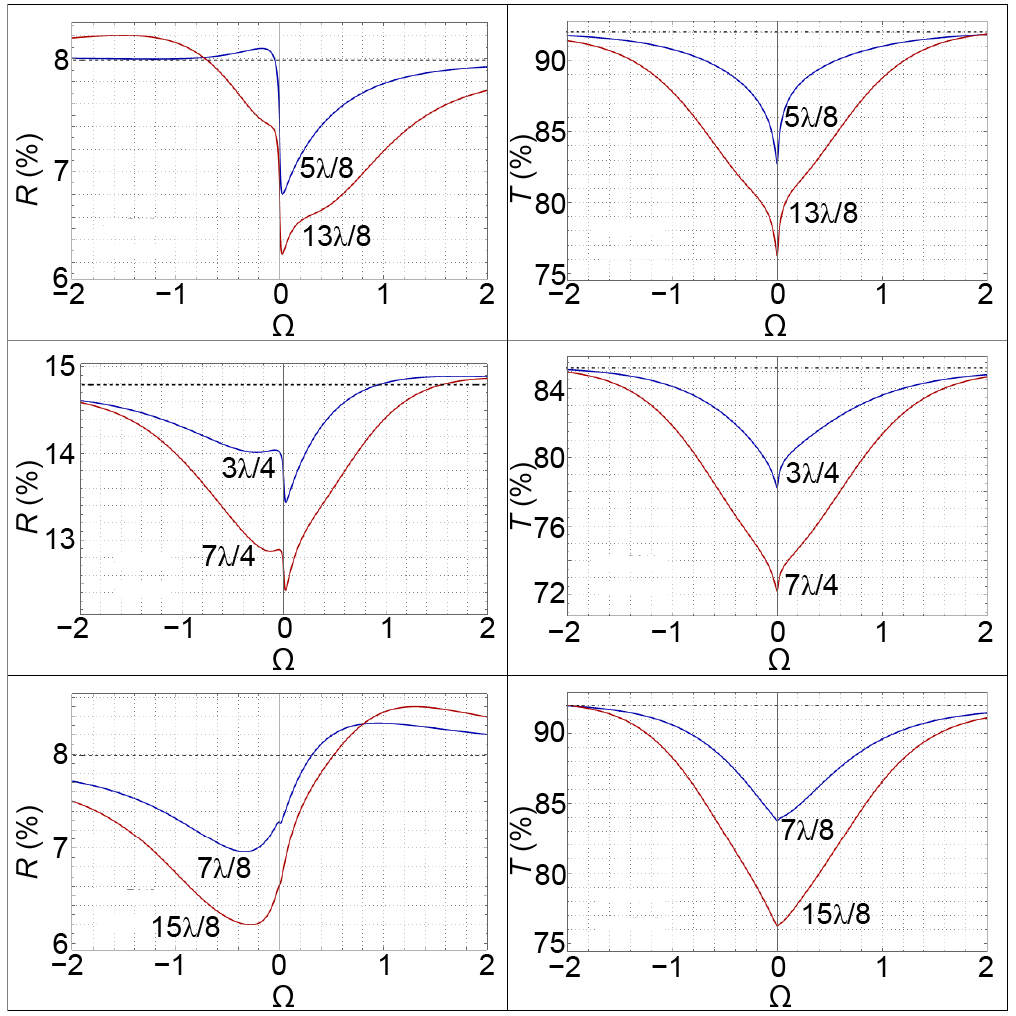}
\caption{\label{fig: periodicity}Reflectivity (left column) and transmittivity (right column) of the vapor layer at various thicknesses calculated for $\gamma/k\upsilon_{T}=0.01$, $m=0.002$, and $n_{1} = n_{2} = 1.5$ in the first order of PT. The observed pattern underlines the $\lambda$-periodicity of the sub-Doppler contribution resulting from the vapor. The horizontal dashed (in reflection) and dot-dashed (in transmission) black lines correspond to the spectra of the empty FP resonator. The error of the calculations is of the order of $m^2$.}
\end{figure}

\begin{figure}[!htb]
\includegraphics{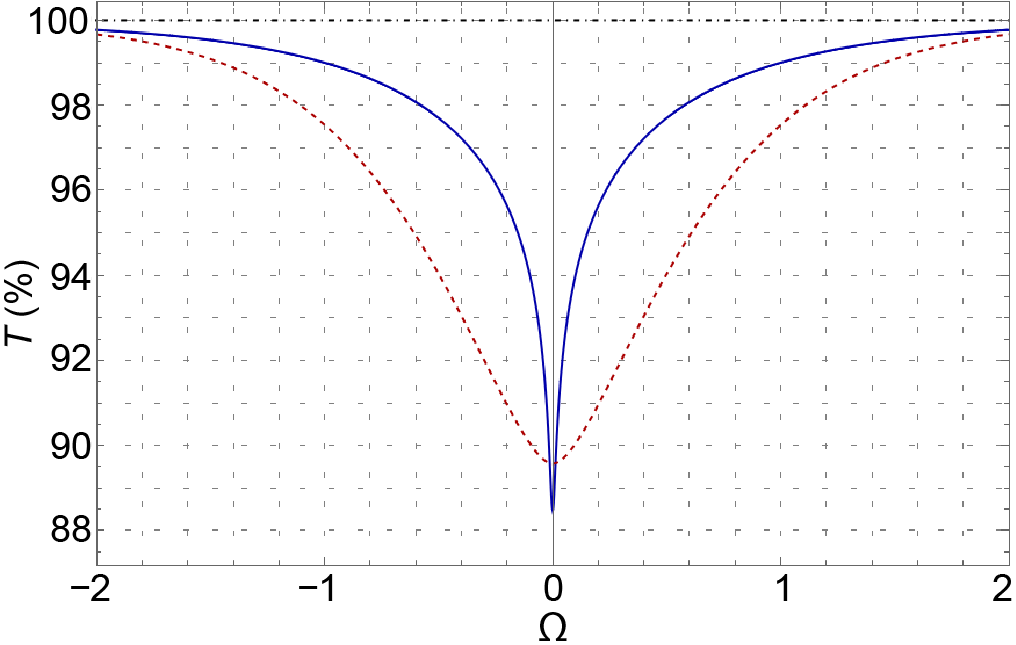}
\caption{Transmittivity of the TVL for two vapor layer thicknesses: $l = \lambda/2$ (blue solid line) and $l = \lambda$ (red dashed line) calculated in the first order of PT for $\gamma/k\upsilon_{T}=0.01$, $m=0.002$ and $n_{1}=n_{2}=1.5$. In both cases transmission of the empty FP resonator is 100\%. The error of the calculations is of the order of $m^2$.}
\label{fig: trans}
\end{figure}
Indeed, as it was first demonstrated in Ref.~\cite{PhysRevA.51.1959} (there the FP effect was not accounted for due to the assumed presence of an antireflection coating on the rear window) the greatest narrowing of spectral lines occurs at a half-integer thicknesses of the gas layer with respect to the wavelength of the incident light. It was also shown that the spectral line shape of SR from and transmission through the TVL experience a $\lambda$-periodic dependence on the gas layer thickness in contrast to the ordinary $\lambda/2$-periodic oscillations that result from the FP cavity effect. Similar effects were previously observed experimentally in the work of Dicke in the radiofrequency domain~\cite{DICKE}. Spectra of reflection and transmission of TVL obtained within the scope of first-order PT were later described in many works as the manifestation of the transient nature of atomic polarization induced by the atom-wall collisions. In Ref.~\cite{Dutier:03} it was pointed out that in the first-order vapor density solution, the mixing of selective contribution of atomic vapor with the FP cavity effect leads to zero reflection from TVL at $l = (2n+1)\lambda/2$, what is consistent with the result presented in Fig.~\ref{fig:first order}. In Fig.~\ref{fig: periodicity} we illustrate also mentioned above $\lambda$-periodic dependence of spectra on the gas layer thickness. It is important to highlight that with the increase in the layer thickness, the sub-Doppler features start to be masked under the broad Doppler spectral line contour arising due to the absorption inside the vapor. In fact, for thicknesses equal to an integer number of wavelengths, the phase mixing of the contributions coming from the different parts of the cavity result in the Doppler-broadened spectral line contour~\cite{DICKE}. To illustrate this, in Fig.~\ref{fig: trans} we present the comparison of spectral line contours in transmission for $l=\lambda/2$ and $l = \lambda$. Indeed, at $l = \lambda$ we observe a purely Doppler spectral line contour in contrast to the sharp sub-Doppler feature at the half-wave thickness. These circumstances underlie the selection of the most attractive range of vapor layer thicknesses $l \sim \lambda/2$ in the context of the considered problem. For more details on spectral peculiarities obtained with the first-order PT solution in optics see~\cite{PhysRevA.51.1959,ZAMBON1997308,Dutier:03}.

We also would like to note here for the first time that in the first-order of PT at any thickness the absorptivity of the vapor layer manifests an even spectral line contour with respect to the atomic transition frequency with the maximum occurring at zero detuning (see Fig.\ref{fig:first order}). This fact could be also verified analytically by substituting the first-order expressions for $r$ and $t$ [See Eqs.~(\ref{eq:r}),~(\ref{eq:t}), and~(\ref{eq:I11})-(\ref{eq:I44})] into the relation Eq.~(\ref{eq:A}). In what follows we demonstrate that the observed symmetry in the absorption spectral line shape is, in fact, the artefact of the first-order PT solution. 

\subsection{Second order of PT}
An undoubted advantage of the first-order solution presented in many works is the relative simplicity of calculating the reflectance and transmittance of TVL. However, the first-order approximation does not correctly account for the effects of light absorption inside the layer and the blueshift of the resonant frequency. Below we present for the first time the second-order solution of the TVL problem for the case of quenching atom-wall collisions. For this purpose, we substitute obtained first-order field solutions [see Eqs.~(\ref{eq:EVEN1SOL})-(\ref{eq:ODD1SOL})] into the iterative equation~(\ref{eq:GREENl})
\begin{eqnarray}
E_{e/o}^{(2)}\left( \xi  \right)=\int_{0}^{\infty } d\nu \int_{{\phi }/{2}}^{\xi }{d{\xi }'\sin \left( {\xi }-{\xi}'  \right)} \\ \nonumber \times \int_{0}^{\phi }{E_{e/0}^{(1)} \left({\xi }'' \right) K\left( \left| {\xi }'-{\xi }'' \right|,\nu \right)d{\xi }''}.
\label{eq:EVENODD2}
\end{eqnarray}
Similarly, the spatial integration can be performed analytically, however the structure of the solution is more complicated. In fact, the final solution in the second order could be obtained by means of two-dimensional velocity integration.
\begin{figure*}[!htb]
\centering
   \includegraphics{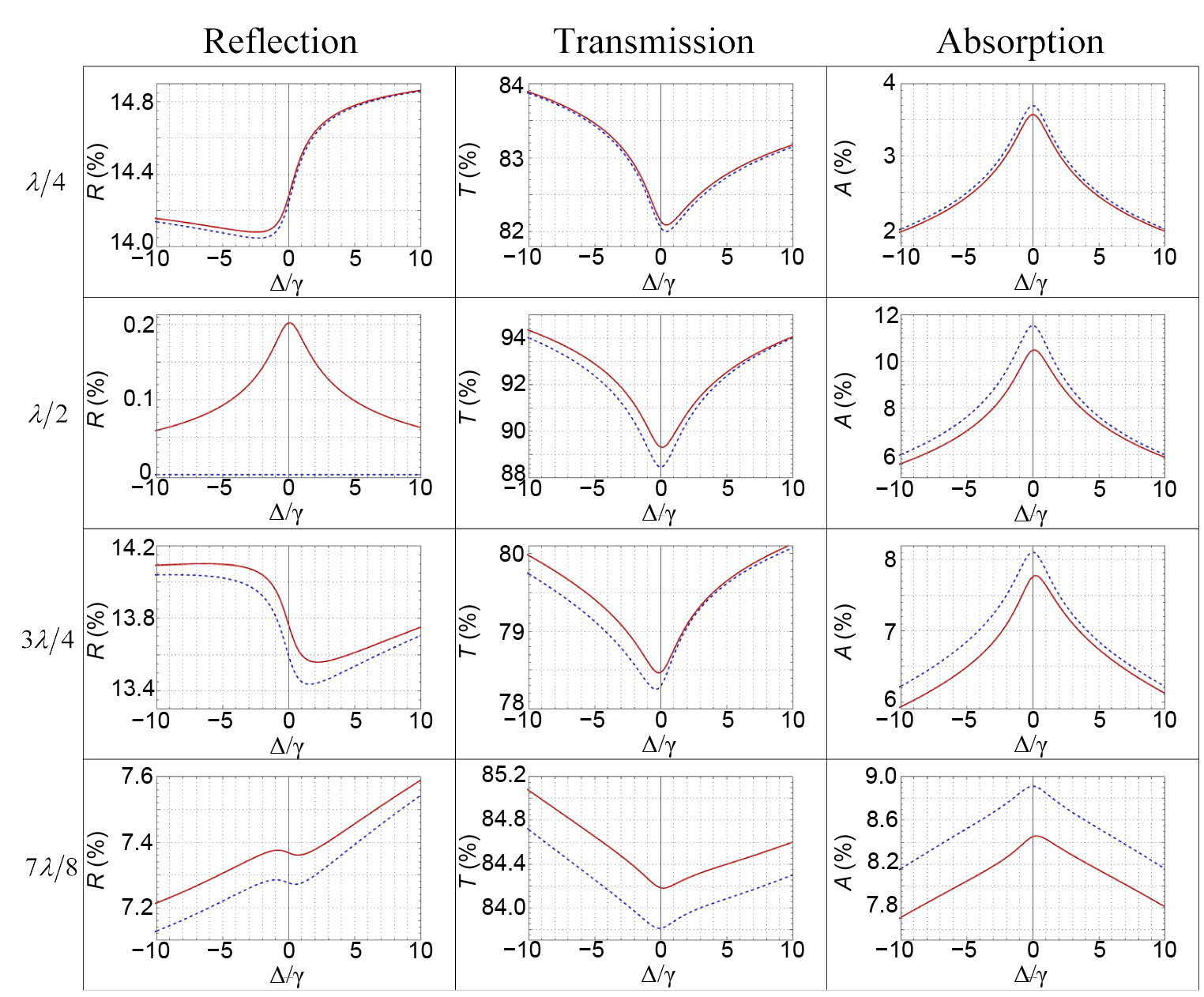}
    \caption{Comparison of reflection, transmission and absorption spectra of the TVL sandwiched between two transparent dielectric media with $n_{1} = n_{2} = 1.5$ calculated up to the first (blue dashed curves) and second (red solid curves) orders of PT in the vicinity of the atomic transition frequency. The calculations were performed using Eqs.~(\ref{eq:EVEN1}),~(\ref{eq:EVENODD2}),~(\ref{eq:r}), and~(\ref{eq:t}) expanded to the second order in $m$. We took the following parameters for our computations: $\gamma/k\upsilon_{T}=0.01$, $m=0.002$. Note that contrary to all previous graphs in these spectra detuning is divided by homogeneous width $\gamma$ rather than Doppler width.} 
    \label{Fig: second order}
\end{figure*}
In Fig.~\ref{Fig: second order} we present the comparison of spectral line shape of reflection, transmission and absorption calculated up to the first and second orders of the PT. At first glance, we can see that the second order contributions with respect to $m$ lead to the change in the amplitude of the sub-Doppler structure, deformation of the spectral line contours and to the noticeable blueshift. At $l = \lambda/2$ in reflection we observe a weak sub-Doppler resonance purely resulting from the second order correction in vapor density. Moreover, in this particular case, the transmission and absorption spectra do not exhibit a symmetrical profile of the spectral line (in contrast to the calculations performed in the first-order of PT), but undergo a blue shift with respect to the atomic transition frequency. In fact, the blueshift manifests itself in all the presented in Fig.~\ref{Fig: second order} transmission and absorption spectra, whereas its value is found to be a complicated function of the number density, thickness, and the width of the spectral line. It is also important to emphasize that the discrepancy between the calculated in the first and second orders of PT becomes more noticeable with an increase in the thickness of the gas layer, i.e. with the increasing influence of the absorption effect.
Finally, we could see from above that the second order correction may significantly modify the spectral line shape of selective reflection, transmission and absorption. This effect could become especially noticeable on experiments at the moderate concentrations of atomic vapor confined inside the spectroscopic nanocell with the thickness $l \sim \lambda/2$ when the widths of the spectra reach their minima of the order of $\gamma$.

\section{\label{sec:Discussion}Discussion and Conclusion}
Linear optical properties of ultrathin layers of rarefied atomic vapors turn out to be very sensitive to the variation of vapor density and to the nature of atom-wall interactions. We have revealed that even in the low vapor concentration limit $m = 2\sqrt{\pi}Nd^{2}/\hbar k \upsilon_{T} \ll 1$ higher-order effects in optical density become significant. In the framework of this research for the first time we have introduced the general approach for constructing the eigenmodes of the TVL beyond the limits of the first-order PT for the case of quenching atom-wall collisions. In particular, it allowed us to demonstrate that the second-order optical density corrections lead to the noticeable deformation of spectral line shapes, to a significant change in the amplitude of the Doppler-free contours, and result in the peculiar blue shift of the resonance frequency comparable to the homogeneous width of the spectral lines [see Fig.~\ref{Fig: second order}]. These results are especially relevant for the implementation of miniaturized atomic sensors based on hot atomic vapor, which usually requires precise control of spectral line widths and the resonance frequency shifts. It is also worth noting that the presented PT approach allows to determine unique solutions for the field inside the vapor layer regardless of the optical properties of the surrounding media. Indeed, the eigenmodes of the TVL can be calculated using the Eqs.~(\ref{eq:series})-(\ref{eq:iteration}) without additional assumptions about the field structure within the layer, which are usually associated with the imposition of dielectric boundary conditions at the initial stage of the calculations. This achievement of the proposed approach is promising in the development of devices based on the resonant interaction of light with plasmonic nanostructures surrounded by alkali metal vapors~\cite{Uriel,URIEL:21}, and prospective in the implementation of gas cells with temperature tunable parameters~\cite{Papoyan1,Papoyan2}.

Ii is important to point out the similarities and differences appearing in the TVL spectra in the case of quenching atom-wall interactions considered in this paper and under the assumption of specular atom-wall collisions, for which a rigorous solution of the self-consistent field problem was already found~\cite{PhysRevA.101.053850}. First of all, for both types of boundary conditions it has been demonstrated that higher order corrections to the optical density lead to a significant deformation of the spectral line shape of the Doppler-free structures. The most remarkable feature of the spectra obtained for the specular boundary conditions is the appearance of a large Lorentzian contribution to the reflection in the vicinity of the resonant frequency, most noticeable at layer thicknesses $l = (2n+1)\lambda/2$, where $n$ is an integer. To our understanding, this peculiarity arises due to the accumulation of polarization induced by the external field by velocity groups of atoms bouncing between two closely spaced dielectric walls without quenching of electron excitation. This argument is also supported by the sharp decrease in the amplitude of the Lorentzian spectral line contour contribution with increasing gas layer thickness, and by the absence of the similar effect in the case of quenching atom-wall collisions. 

Another notable feature of the TVL spectra calculated beyond the scope of the first-order PT is the presence of the large blueshift for both models of atom-wall interactions. This shift of the purely electromagnetic nature was previously studied in the case of optically-thick gas layer (see, for example,~\cite{schuurmans:jpa-00208442,GUO1996}), where it was attributed to the mentioned above transient process of establishing polarization followed by the atom-wall collisions. A more detailed examination of this effect showed that the blueshift in reflection from a semi-infinite vapor layer is largely determined by the interference process of contributions from two classes of atoms: "arriving" and "departing" from the glass-vapor interface~\cite{VART2001}. To accurately consider the dependence of the blueshift on the concentration of atomic vapors and other parameters of the system, the Maxwell-Bloch set of equations in the medium should be solved self-consistently. Under the assumption of specular atom-wall collisions we found the following linear proportionality of the blueshift in reflection spectra for the most interesting case of half-integer layer thickness
\begin{eqnarray}
\Delta_{s}(l = \lambda/2) = 3.54m,
\label{eq:shift_spec}
\end{eqnarray}
where $m \leq \Gamma$ is limited by the self-broadening effect in the strong spatial dispersion domain $\Gamma = \gamma/k\upsilon_{T} = 0.01 \ll 1$, subscript "$s$" indicates that the universal proportionality constant was found for the case of specular boundary conditions. 

In the case of quenching collisions, the analysis of the blueshift in the reflection proves to be more difficult, since the Doppler-free structures in the reflection are no longer exhibit even behavior with respect to the frequency detuning (see Fig.~\ref{Fig: second order}). In fact, the shift of the resonance frequency is a well-defined parameter only for an even spectral line contour with one maximum. In Sec.~\ref{sec:Zero}, we pointed out a remarkable feature of the spectral line shapes of absorption bands calculated in the first-order of PT, namely, such spectral line contours are even with respect to the resonance transition with the maximum at $\Omega = 0$ for any thickness of the gas layer. This circumstance turns out to be useful in studying the effect of higher-orders on the density-dependent shift of spectral lines at different thicknesses of the gas layer. Below we provide the obtained dependencies of the blueshift in absorption on the optical density of atomic vapor for two vapor layer thicknesses $l = \lambda/2$ and $l = 3\lambda/4$, at which the blueshift is large enough, while the spectral line width remains sub-Doppler. The calculations performed in the second order of PT show that these dependencies can be approximated with a sufficient accuracy by linear functions (the error of these calculations is of the order of $m^3$)
\begin{eqnarray}
\Delta_{q}(l = \lambda/2) = 0.56m, \\
\Delta_{q}(l = 3\lambda/4) = 1.21m,
\label{eq:shift_diff}
\end{eqnarray}
where subscript "$q$" stands for quenching atom-wall collisions. It can be directly seen that the blueshift of the resonance frequency has values comparable to the homogeneous width of the spectral lines ($\Gamma = 0.01$) for two considered models of atom-wall interactions. This result is especially important with respect to the experiment and practical applications, since implies that this blueshift phenomenon should be taken into account in any attempts of studying a wide class of effects leading to the deformation of spectral line shape and shift of the resonance frequency. A detailed study of the structure of the second-order corrections shows that such large values of the blueshift arise from a remarkable term proportional to $\eta^{-1}$, which was absent in the first order solution. 

In conclusion, we would like to emphasize that the results obtained in this paper are an important step towards a fundamental understanding of the emergence and interplay of numerous processes of atom-wall and atom-light interactions at nanoscale, which prevail in nanocells filled with hot atomic vapor. Needless to say, the constructed theory does not exhaust the whole variety of these complex processes. However, we have consistently demonstrated that the spectra of reflection, transmission and absorption of TVL turn out to be significantly dependent on the higher-orders optical density effects, which have not previously been taken into account for the case of quenching atomic-wall collisions. Speaking about the prospects in the field of ultra-thin vapor cell spectroscopy, it is of great interest to construct a rigorous solution to the TVL problem with diffuse boundary conditions (similar to the one we obtained earlier for the case of specular boundary conditions~\cite{PhysRevA.101.053850} and another one obtained by Schuurmans for the thick gas layer case earlier~\cite{schuurmans:jpa-00208442}). By analogy with the problem of the anomalous skin effect~\cite{LIFSHITZ1981329}, such a solution can be obtained via the Wiener-Hopf method repeatedly used to solve similar physical problems (see, for example,~\cite{Baraff}). An exact solution to this problem will allow to further investigate the complex dependence of the shift and broadening of the spectral lines of the TVL reflection and transmission on the system parameters.

\begin{widetext}
\appendix*
\section{First order of PT}
\label{sec: 1st ORDER}
In the first order of PT, the exact expressions for $I_{1} - I_{4}$ read as
\begin{eqnarray}
I_{1}^{(1)}=\cos \left( \frac{\phi }{2} \right)-i m \phi \sin \left( \frac{\phi }{2} \right)\int_{0}^{\infty } d\nu \frac{\eta {{\text{e}}^{-{{\nu }^{2}}}}}{{{\eta }^{2}}+{{\nu }^{2}}} \nonumber \\ -2 i m\int_{0}^{\infty } d\nu \frac{{{\nu }^{2}}{{\text{e}}^{-{{\nu }^{2}}}}}{{{\left( {{\eta }^{2}}+{{\nu }^{2}} \right)}^{2}}}{{e}^{-\frac{\phi \eta }{\nu }}}\left[ 1+{{e}^{\frac{\phi \eta }{\nu }}}-2{{e}^{\frac{\phi \eta }{2\nu }}}\cos \left( \frac{\phi }{2} \right) \right] \left[ \nu \sin \left( \frac{\phi }{2} \right)-\cos \left( \frac{\phi }{2} \right)\eta  \right],
\label{eq:I11}
\end{eqnarray}

\begin{eqnarray}
I_{2}^{(1)}=-\sin \left( \frac{\phi }{2} \right) - i m \phi \cos \left( \frac{\phi }{2} \right)\int_{0}^{\infty } d\nu \frac{\eta {{e}^{-{{\nu }^{2}}}}}{{{\eta }^{2}}+ {{\nu }^{2}}} \nonumber \\ +2 i m\int_{0}^{\infty} d\nu\frac{{{e}^{-{{\nu }^{2}}}}}{{{\left( {{\eta }^{2}}+{{\nu }^{2}} \right)}^{2}}} \Bigg\{ -{{\nu }^{3}}\cos \left( \frac{\phi }{2} \right)+\sin \left( \frac{\phi }{2} \right){{\eta }^{3}} \nonumber \\ +{{e}^{-\frac{\phi \eta }{2\nu }}}\nu \eta \left( \nu \sin \phi +\eta -\eta \cos \phi  \right)+{{e}^{-\frac{\phi \eta }{\nu }}}{{\nu }^{2}} \left[ \nu \cos \left( \frac{\phi }{2} \right)+\sin \left( \frac{\phi }{2} \right)\eta  \right] \Bigg\},
\label{eq:I22}
\end{eqnarray}

\begin{eqnarray}
I_{3}^{(1)}=\sin \left( \frac{\phi }{2} \right) + i m \phi \cos \left( \frac{\phi }{2} \right)\int_{0}^{\infty } d\nu \frac{\eta {{e}^{-{{\nu }^{2}}}}}{{{\eta }^{2}}+{{\nu }^{2}}} \nonumber \\ + 2 i m\int_{0}^{\infty } d\nu \frac{{{e}^{-{{\nu }^{2}}}}}{{{\left( {{\eta }^{2}}+{{\nu }^{2}} \right)}^{2}}} \Bigg\{ -\nu \cos \left( \frac{\phi }{2} \right){{\eta }^{2}}+{{e}^{-\frac{\phi \eta }{\nu }}}\nu \eta \left[ -\nu \sin \left( \frac{\phi }{2} \right)+\cos \eta  \right] \nonumber \\ -{{e}^{-\frac{\phi \eta }{2\nu }}}{{\nu }^{2}} \left( \nu \left[ -1+\cos \phi  \right]+\eta \sin \phi  \right)+\sin \left( \frac{\phi }{2} \right)\eta \left( 2{{\nu }^{2}}+{{\eta }^{2}} \right) \Bigg\},
\label{eq:I33}
\end{eqnarray}

\begin{eqnarray}
I_{4}^{(1)}=\cos \left( \frac{\phi }{2} \right) - i m \phi \sin \left( \frac{\phi }{2} \right)\int_{0}^{\infty } d\nu \frac{\eta {{e}^{-{{\nu }^{2}}}}}{{{\eta }^{2}}+{{\nu }^{2}}} \nonumber \\ + 2 i m\int_{0}^{\infty } d\nu\frac{\nu \eta {{e}^{-{{\nu }^{2}}}}}{{{\left( {{\eta }^{2}}+{{\nu }^{2}} \right)}^{2}}}{{e}^{-\frac{\phi \eta }{\nu }}}\left[ 1+{{e}^{\frac{\phi \eta }{\nu }}}-2{{e}^{\frac{\phi \eta }{2\nu }}}\cos \left( \frac{\phi }{2} \right) \right] \left[ \nu \cos \left( \frac{\phi }{2} \right)+\eta \sin \left( \frac{\phi }{2} \right) \right].
\label{eq:I44}
\end{eqnarray}
\end{widetext}

The structure of Eqs.~(\ref{eq:I11})-(\ref{eq:I44}) is the following: the first terms are independent of the atomic number density and come from the solution for the field in a vacuum, in other words, they represent the zero-order optical density solution. Together these terms in expressions above result in the usual $\lambda/2$-periodic dependence of the reflectivity and transmittivity on the layer thickness arising due to the FP effect. The middle terms in the above expressions originate from the steady-state polarization component and give a rise to a wide Doppler spectral line contour. At first glance it is peculiar that this particular terms are proportional not only to atomic number density, but also to the gas layer thickness $\phi$. In fact, this result is just the first term of the series expansion of the conventional exponential absorption pattern and consequently this limits the largest gas layer thickness accessible in the first order of PT. The most interesting part of expressions~(\ref{eq:I11})-(\ref{eq:I44}) is the last integral terms, which result from the transient behavior of the polarization followed by the atom-wall collisions. In these terms, the dependence on $\phi$ is already included in the exponential terms inside the integral kernel. This feature leads to a $\lambda$-periodic spectral dependence of reflectivity and transmittivity on the layer thickness [see Fig.~\ref{fig: periodicity}]. Moreover, it is this particular contribution that leads to the formation of a Doppler-free structure in the spectra. The explicit form of the above expressions gives us the visual representation of the emerging spectral structures. In fact, after substitution of Eqs.~(\ref{eq:I11})-(\ref{eq:I44}) in Eqs.~(\ref{eq:r})-(\ref{eq:t}), the spectral contours of reflection, transmission, and absorption are formed due to the natural mixing of the FP, Doppler, and the sub-Doppler contributions.

\providecommand{\noopsort}[1]{}\providecommand{\singleletter}[1]{#1}%

\nocite{*}
\end{document}